\documentclass[a4paper,11pt]{article}
\usepackage[latin1]{inputenc}
\usepackage{amssymb}

\usepackage{times}

\title{{\bf   The  Determinacy  of Context-Free Games }} 
\author{Olivier Finkel \\
{\it Equipe de Logique Math\'ematique}
\\Institut de Math\'ematiques de Jussieu
 \\  CNRS et Universit\'e Paris 7, France. \\ 
finkel@logique.jussieu.fr }

\date{}
\begin{document}

\newtheorem{theorem}{Theorem}[section]
\newtheorem{Rem}[theorem]{Remark}
\newtheorem{Exa}[theorem]{Example}

\newtheorem{Pro}[theorem]{Proposition}
\newtheorem{lem}[theorem]{Lemma}
\newtheorem{Cor}[theorem]{Corollary}
\newtheorem{defi}[theorem]{Definition}
\newtheorem{notation}[theorem]{Notation}

\def\ufootnote#1{\let\savedthfn\thefootnote\let\thefootnote\relax
\footnote{#1}\let\thefootnote\savedthfn\addtocounter{footnote}{-1}}

\newcommand{\bormxi}{{\bf\Pi}^{0}_{\xi}}
\newcommand{\bormlxi}{{\bf\Pi}^{0}_{<\xi}}
\newcommand{\bormz}{{\bf\Pi}^{0}_{0}}
\newcommand{\bormone}{{\bf\Pi}^{0}_{1}}
\newcommand{\ca}{{\bf\Pi}^{1}_{1}}
\newcommand{\bormtwo}{{\bf\Pi}^{0}_{2}}
\newcommand{\bormthree}{{\bf\Pi}^{0}_{3}}
\newcommand{\bormom}{{\bf\Pi}^{0}_{\omega}}
\newcommand{\borom}{{\bf\Delta}^{0}_{\omega}}
\newcommand{\borml}{{\bf\Pi}^{0}_{\lambda}}
\newcommand{\bormlpn}{{\bf\Pi}^{0}_{\lambda +n}}
\newcommand{\bormpm}{{\bf\Pi}^{0}_{1+m}}
\newcommand{\borapm}{{\bf\Sigma}^{0}_{1+m}}
\newcommand{\bormep}{{\bf\Pi}^{0}_{\eta +1}}
\newcommand{\borapxi}{{\bf\Sigma}^{0}_{\xi}}
\newcommand{\borai}{{\bf\Sigma}^{0}_{ 2.\xi +1 }}
\newcommand{\bormpxi}{{\bf\Pi}^{0}_{\xi}}
\newcommand{\bormpeta}{{\bf\Pi}^{0}_{1+\eta}}
\newcommand{\borapxipo}{{\bf\Sigma}^{0}_{\xi +1}}
\newcommand{\bormpxipo}{{\bf\Pi}^{0}_{\xi +1}}
\newcommand{\borpxi}{{\bf\Delta}^{0}_{\xi}}
\newcommand{\borel}{{\bf\Delta}^{1}_{1}}
\newcommand{\Borel}{{\it\Delta}^{1}_{1}}
\newcommand{\borone}{{\bf\Delta}^{0}_{1}}
\newcommand{\bortwo}{{\bf\Delta}^{0}_{2}}
\newcommand{\borthree}{{\bf\Delta}^{0}_{3}}
\newcommand{\boraone}{{\bf\Sigma}^{0}_{1}}
\newcommand{\boratwo}{{\bf\Sigma}^{0}_{2}}
\newcommand{\borathree}{{\bf\Sigma}^{0}_{3}}
\newcommand{\boraom}{{\bf\Sigma}^{0}_{\omega}}
\newcommand{\boraxi}{{\bf\Sigma}^{0}_{\xi}}
\newcommand{\ana}{{\bf\Sigma}^{1}_{1}}
\newcommand{\pca}{{\bf\Sigma}^{1}_{2}}
\newcommand{\Ana}{{\it\Sigma}^{1}_{1}}
\newcommand{\Boraone}{{\it\Sigma}^{0}_{1}}
\newcommand{\Borone}{{\it\Delta}^{0}_{1}}
\newcommand{\Bormone}{{\it\Pi}^{0}_{1}}
\newcommand{\Bormtwo}{{\it\Pi}^{0}_{2}}
\newcommand{\Ca}{{\it\Pi}^{1}_{1}}
\newcommand{\bormn}{{\bf\Pi}^{0}_{n}}
\newcommand{\bormm}{{\bf\Pi}^{0}_{m}}
\newcommand{\boralp}{{\bf\Sigma}^{0}_{\lambda +1}}
\newcommand{\borat}{{\bf\Sigma}^{0}_{|\theta |}}
\newcommand{\bormat}{{\bf\Pi}^{0}_{|\theta |}}
\newcommand{\Borapxi}{{\it\Sigma}^{0}_{\xi}}
\newcommand{\Bormpxipo}{{\it\Pi}^{0}_{1+\xi +1}}
\newcommand{\Borapn}{{\it\Sigma}^{0}_{1+n}}
\newcommand{\borapn}{{\bf\Sigma}^{0}_{1+n}}
\newcommand{\boraxipm}{{\bf\Sigma}^{0}_{\xi^\pm}}
\newcommand{\Boratwo}{{\it\Sigma}^{0}_{2}}
\newcommand{\Borathree}{{\it\Sigma}^{0}_{3}}
\newcommand{\Borapnpo}{{\it\Sigma}^{0}_{1+n+1}}
\newcommand{\Bormpxi}{{\it\Pi}^{0}_{\xi}}
\newcommand{\Borpxi}{{\it\Delta}^{0}_{\xi}}
\newcommand{\boratpxi}{{\bf\Sigma}^{0}_{2+\xi}}
\newcommand{\Boratpxi}{{\it\Sigma}^{0}_{2+\xi}}
\newcommand{\bormltpxi}{{\bf\Pi}^{0}_{<2+\xi}}
\newcommand{\Bormltpxi}{{\it\Pi}^{0}_{<2+\xi}}
\newcommand{\borapeap}{{\bf\Sigma}^{0}_{1+\eta_{\alpha ,p}}}
\newcommand{\borapeapn}{{\bf\Sigma}^{0}_{1+\eta_{\alpha ,p,n}}}
\newcommand{\Borapeap}{{\it\Sigma}^{0}_{1+\eta_{\alpha ,p}}}
\newcommand{\Bormpn}{{\it\Pi}^{0}_{1+n}}
\newcommand{\Borpn}{{\it\Delta}^{0}_{1+n}}
\newcommand{\borapximo}{{\bf\Sigma}^{0}_{1+(\xi -1)}}
\newcommand{\borpeta}{{\bf\Delta}^{0}_{1+\eta}}

\newcommand{\hs}{\hspace{12mm}

}
\newcommand{\noi}{\noindent}

\newcommand{\om}{\omega}
\newcommand{\Si}{\Sigma}
\newcommand{\Sis}{\Sigma^\star}
\newcommand{\Sio}{\Sigma^\omega}
\newcommand{\nl}{\newline}
\newcommand{\lra}{\leftrightarrow}
\newcommand{\fa}{\forall}
\newcommand{\ra}{\rightarrow}
\newcommand{\orl}{ $\omega$-regular language}

\newcommand{\Ga}{\Gamma}
\newcommand{\Gas}{\Gamma^\star}
\newcommand{\Gao}{\Gamma^\omega}
\newcommand{\ite}{\item}
\newcommand{\la}{language}
\newcommand{\Lp}{L(\varphi)}
\newcommand{\abs}{\{a, b\}^\star}
\newcommand{\abcs}{\{a, b, c \}^\star}
\newcommand{\ol}{$\omega$-language}

\newcommand{\tla}{\twoheadleftarrow}
\newcommand{\de}{deterministic }
\newcommand{\proo}{\noi {\bf Proof.} }
\newcommand {\ep}{\hfill $\square$}

\maketitle

\begin{abstract}
\noi We prove that  the determinacy of   Gale-Stewart games whose winning sets are accepted by 
real-time  $1$-counter B\"uchi automata  is  equivalent to the determinacy of (effective) analytic Gale-Stewart games which is known to be a large cardinal 
assumption. We show also that the determinacy of Wadge games between two players in charge of 
$\om$-languages accepted by  $1$-counter B\"uchi automata  is  equivalent to the  (effective) analytic  Wadge determinacy. 
Using some results of set theory we prove that  one can effectively construct a 
 $1$-counter B\"uchi automaton $\mathcal{A}$ and a B\"uchi automaton $\mathcal{B}$ such that: (1) There exists a model of  ZFC 
in which Player 2 has a winning strategy in the Wadge game $W(L(\mathcal{A}), L(\mathcal{B}))$; (2)  There exists a model of  ZFC 
in which the Wadge game $W(L(\mathcal{A}), L(\mathcal{B}))$ is not determined. 
Moreover these are the only two possibilities, i.e. there are no models of ZFC in which 
Player 1  has a 
winning strategy in the Wadge game $W(L(\mathcal{A}), L(\mathcal{B}))$.
\end{abstract}


\section{Introduction}

Two-players infinite games have been much studied in Set Theory and  in Descriptive Set Theory, see \cite{Kechris94,Jech}. In particular, if $X$ is a (countable) alphabet having at 
least two letters and $A \subseteq X^\om$, then the Gale-Stewart game $G(A)$ is an infinite  game with perfect
 information between two players. Player 1 first writes a letter
$a_1\in X$, then Player 2 writes a letter $b_1\in X$,
 then Player 1 writes $a_2\in X$, and so on $\ldots$
After $\om$ steps, the two players have composed an infinite word 
$x =a_1b_1a_2b_2\ldots$ of $X^\om$.
 Player 1 wins the play iff $x \in A$, otherwise Player 2
wins the play. The game $G(A)$ is said to be determined iff 
one of the two players has a winning strategy. A fundamental result of Descriptive Set Theory is 
Martin's Theorem which states that every Gale-Stewart game
 $G(A)$, where $A$ is a Borel set, is determined \cite{Kechris94}. 

On the other hand, in Computer Science, the conditions
 of a Gale Stewart game may be seen as a specification of a reactive system,
 where the two players are respectively  a non terminating reactive
 program and  the  ``environment".
 Then the problem of the synthesis of winning strategies
is of great practical
interest for the problem of program synthesis in reactive systems.
In particular, if $A \subseteq X^\om$, where $X$ is here a finite alphabet, and $A$ is effectively presented, i.e. accepted by a given 
finite machine or defined by a given logical formula, the following questions naturally arise, see  \cite{Thomas95,LescowThomas}:
(1) ~ Is the game $G(A)$ determined ?
~ (2) ~ If Player 1 has a winning strategy, is it effective, i.e. computable ? 
~(3) ~ What are the amounts of space and time necessary to compute such a winning strategy ? 
B\"uchi and Landweber  gave a solution to the famous Church's Problem, posed in 1957,   by stating  that in a Gale Stewart game $G(A)$,
where $A$ is a regular $\om$-language, one can decide who is the winner and
compute a winning strategy given by a finite state transducer, see \cite{Thomas08} for more information on this 
subject. In \cite{Thomas95,LescowThomas} Thomas and Lescow asked for an extension of this
result where $A$ is no longer regular but deterministic context-free, i.e. accepted by some deterministic pushdown automaton. 
 Walukiewicz  extended B\"uchi and Landweber's Theorem to this case by showing first  in  \cite{wal} that 
that one can effectively construct winning strategies in parity games played on
pushdown graphs and that these strategies can be computed by pushdown
transducers. Notice that later some extensions to the case of higher-order pushdown automata have been established.

In this paper, we first address the question (1) of the determinacy of Gale-Stewart games $G(A)$, 
where $A$ is a context-free $\om$-language accepted by a (non-deterministic)  pushdown automaton, or 
even by a $1$-counter automaton. Notice that  there are some context-free $\om$-languages which are (effective) analytic but non-Borel and thus 
the determinacy of these games can not be deduced from Martin's Theorem of Borel determinacy. On the other hand, Martin's Theorem is provable 
in ZFC, the commonly accepted axiomatic 
framework for Set Theory in which all usual mathematics can be developped.  But  the determinacy of Gale-Stewart games $G(A)$,  where 
$A$ is an (effective) analytic set, is not provable in ZFC; Martin and Harrington have proved that it is  a large cardinal 
assumption equivalent to the existence of a particular real, called the real $0^\sharp$, see \cite[page  637]{Jech}. We prove here that  the determinacy of   
Gale-Stewart games  $G(A)$, whose winning sets $A$ are accepted by 
real-time  $1$-counter B\"uchi automata,  is  equivalent to the determinacy of (effective) analytic Gale-Stewart games and thus also 
 equivalent to the existence of the real $0^\sharp$.

Next we consider Wadge games which were firstly studied      by  Wadge     in \cite{Wadge83} where he determined 
a great refinement of the Borel hierarchy defined 
via  the notion of  reduction by continuous functions.  These games are closely related to the notion of reducibility by continuous  functions.  
For $L\subseteq X^\om$ and $L'\subseteq Y^\om$, $L$ is said to be Wadge reducible to $L'$ 
 iff there exists a continuous function $f: X^\om \ra Y^\om$, such that
$L=f^{-1}(L')$; this is then denoted by $L\leq _W L'$.  On the other hand, the Wadge game $W(L, L')$ is  an infinite  game with perfect information between two players,
Player 1 who is in charge of $L$ and Player 2 who is in charge of $L'$.  And it turned out that Player 2 has a winning strategy in the Wadge game 
 $W(L, L')$  iff  $L\leq _W L'$. It is easy to see that the determinacy of Borel Gale-Stewart games implies the determinacy of Borel Wadge games. On the other
hand,  Louveau and Saint-Raymond have  proved that 
this latter one is weaker than the first one,  since it is already provable in second-order arithmetic,   
while the first one is not. 
 It is also known that the determinacy of (effective) analytic 
Gale-Stewart games is equivalent to the determinacy of (effective) analytic Wadge games,  see \cite{Louveau-Saint-Raymond}. 
We prove in this paper  that the determinacy of Wadge games between two players in charge of 
$\om$-languages accepted by  $1$-counter B\"uchi automata  is  equivalent to the  (effective) analytic  Wadge determinacy, and thus also 
equivalent to the existence of the real $0^\sharp$. 

Then, using some recent results from \cite{Fin-ICST} and some results of Set Theory,  we prove that, (assuming  ZFC is consistent),   one can effectively construct a 
 $1$-counter B\"uchi automaton $\mathcal{A}$ and a B\"uchi automaton $\mathcal{B}$ such that: (1) There exists a model of  ZFC 
in which Player 2 has a winning strategy in the Wadge game $W(L(\mathcal{A}), L(\mathcal{B}))$; (2)  There exists a model of  ZFC 
in which the Wadge game $W(L(\mathcal{A}), L(\mathcal{B}))$ is not determined. 
Moreover these are the only two possibilities, i.e. there are no models of ZFC in which 
Player 1  has a 
winning strategy in the Wadge game $W(L(\mathcal{A}), L(\mathcal{B}))$.

The paper is organized as follows. We recall some known notions in Section 2. We study context-free Gale-Stewart games in Section 3 and 
context-free Wadge  games  in Section 4. Some concluding remarks are given in Section 5.

\section{Recall of some known notions}
 
~~~~~  We assume   the reader to be familiar with the theory of formal ($\om$-)languages  
\cite{Staiger97,PerrinPin}.
We recall the  usual notations of formal language theory. 

If  $\Si$ is a finite alphabet, a {\it non-empty finite word} over $\Si$ is any 
sequence $x=a_1\ldots a_k$, where $a_i\in\Sigma$ 
for $i=1,\ldots ,k$ , and  $k$ is an integer $\geq 1$. The {\it length}
 of $x$ is $k$, denoted by $|x|$.
 The {\it empty word} has no letter and is denoted by $\lambda$; its length is $0$. 
 $\Sis$  is the {\it set of finite words} (including the empty word) over $\Sigma$.
A  (finitary) {\it language} $V$ over an alphabet $\Sigma$ is a subset of  $\Sis$.

 The {\it first infinite ordinal} is $\om$.
 An $\om$-{\it word} over $\Si$ is an $\om$ -sequence $a_1 \ldots a_n \ldots$, where for all 
integers $ i\geq 1$, ~
$a_i \in\Sigma$.  When $\sigma=a_1 \ldots a_n \ldots$ is an $\om$-word over $\Si$, we write
 $\sigma(n)=a_n$,   $\sigma[n]=\sigma(1)\sigma(2)\ldots \sigma(n)$  for all $n\geq 1$ and $\sigma[0]=\lambda$.

 The usual concatenation product of two finite words $u$ and $v$ is 
denoted $u.v$ (and sometimes just $uv$). This product is extended to the product of a 
finite word $u$ and an $\om$-word $v$: the infinite word $u.v$ is then the $\om$-word such that:

 $(u.v)(k)=u(k)$  if $k\leq |u|$ , and 
 $(u.v)(k)=v(k-|u|)$  if $k>|u|$.
  
 The {\it set of } $\om$-{\it words} over  the alphabet $\Si$ is denoted by $\Si^\om$.
An  $\om$-{\it language} $V$ over an alphabet $\Sigma$ is a subset of  $\Si^\om$, and its  complement (in $\Sio$) 
 is $\Sio - V$, denoted $V^-$.

  The {\it prefix relation} is denoted $\sqsubseteq$: a finite word $u$ is a {\it prefix} 
of a finite word $v$ (respectively,  an infinite word $v$), denoted $u\sqsubseteq v$,  
 if and only if there exists a finite word $w$ 
(respectively,  an infinite word $w$), such that $v=u.w$.  

If  $L$ is a finitary language (respectively,  an $\om$-language) over   the alphabet $\Si$ then the  set 
${\rm Pref}(L)$ of prefixes of elements of $L$ is defined by  ${\rm Pref}(L)=\{ u\in \Sis \mid  \exists v\in L ~~ u \sqsubseteq v \}$.

 We now recall the definition of $k$-counter B\"uchi automata which will be useful in the sequel. 

 Let $k$ be an integer $\geq 1$. 
A  $k$-counter machine has $k$ {\it counters}, each of which containing a  non-negative integer. 
The machine can test whether the content of a given counter is zero or not. 
And transitions depend on the letter read by the machine, the current state of the finite control, and the tests about the values of the counters. 
Notice that in this model some  $\lambda$-transitions are allowed. During these transitions the reading head of the machine does not move to the right, i.e. 
 the machine does not  read any more letter. 

Formally a  $k$-counter machine is a 4-tuple 
$\mathcal{M}$=$(K,\Si,$ $ \Delta, q_0)$,  where $K$ 
is a finite set of states, $\Sigma$ is a finite input alphabet, 
 $q_0\in K$ is the initial state, 
and  $\Delta \subseteq K \times ( \Si \cup \{\lambda\} ) \times \{0, 1\}^k \times K \times \{0, 1, -1\}^k$ is the transition relation. 
The $k$-counter machine $\mathcal{M}$ is said to be {\it real time} iff: 
$\Delta \subseteq K \times
  \Si \times \{0, 1\}^k \times K \times \{0, 1, -1\}^k$, 
 i.e. iff there are no  $\lambda$-transitions. 

If  the machine $\mathcal{M}$ is in state $q$ and 
$c_i \in \mathbf{N}$ is the content of the $i^{th}$ counter 
 $\mathcal{C}$$_i$ then 
the  configuration (or global state)
 of $\mathcal{M}$ is the  $(k+1)$-tuple $(q, c_1, \ldots , c_k)$.

 For $a\in \Si \cup \{\lambda\}$, 
$q, q' \in K$ and $(c_1, \ldots , c_k) \in \mathbf{N}^k$ such 
that $c_j=0$ for $j\in E \subseteq  \{1, \ldots , k\}$ and $c_j >0$ for 
$j\notin E$, if 
$(q, a, i_1, \ldots , i_k, q', j_1, \ldots , j_k) \in \Delta$ where $i_j=0$ for $j\in E$ 
and $i_j=1$ for $j\notin E$, then we write:

~~~~~~~~$a: (q, c_1, \ldots , c_k)\mapsto_{\mathcal{M}} (q', c_1+j_1, \ldots , c_k+j_k)$.

 Thus  the transition relation must obviously satisfy:
 \nl if $(q, a, i_1, \ldots , i_k, q', j_1, \ldots , j_k)  \in    \Delta$ and  $i_m=0$ for 
 some $m\in \{1, \ldots , k\}$  then $j_m=0$ or $j_m=1$ (but $j_m$ may not be equal to $-1$). 
  
Let $\sigma =a_1a_2 \ldots a_n \ldots $ be an $\om$-word over $\Si$. 
An $\om$-sequence of configurations $r=(q_i, c_1^{i}, \ldots c_k^{i})_{i \geq 1}$ is called 
a run of $\mathcal{M}$ on $\sigma$  iff:

(1)  $(q_1, c_1^{1}, \ldots c_k^{1})=(q_0, 0, \ldots, 0)$

(2)   for each $i\geq 1$, there  exists $b_i \in \Si \cup \{\lambda\}$ such that
 $b_i: (q_i, c_1^{i}, \ldots c_k^{i})\mapsto_{\mathcal{M}}  
(q_{i+1},  c_1^{i+1}, \ldots c_k^{i+1})$  
and such that  ~  $a_1a_2\ldots a_n\ldots =b_1b_2\ldots b_n\ldots$

For every such run $r$, $\mathrm{In}(r)$ is the set of all states entered infinitely
 often during $r$.

\begin{defi} A B\"uchi $k$-counter automaton  is a 5-tuple 
$\mathcal{M}$=$(K,\Si,$ $\Delta, q_0, F)$, 
where $ \mathcal{M}'$=$(K,\Si,$  $\Delta, q_0)$
is a $k$-counter machine and $F \subseteq K$ 
is the set of accepting  states.
The \ol~ accepted by $\mathcal{M}$ is:~~ $L(\mathcal{M})$= $\{  \sigma\in\Si^\om \mid \mbox{  there exists a  run r
 of } \mathcal{M} \mbox{ on } \sigma \mbox{  such that } \mathrm{In}(r)
 \cap F \neq \emptyset \}$

\end{defi}

  The class of \ol s accepted by  B\"uchi $k$-counter automata  is  
denoted ${\bf BCL}(k)_\om$.
 The class of \ol s accepted by {\it  real time} B\"uchi $k$-counter automata  will be 
denoted {\bf r}-${\bf BCL}(k)_\om$.
  The class ${\bf BCL}(1)_\om$ is  a strict subclass of the class ${\bf CFL}_\om$ of context free \ol s
accepted by B\"uchi pushdown automata.

\hs  We assume the reader to be familiar with basic notions of topology which
may be found in \cite{Kechris94,LescowThomas,Staiger97,PerrinPin}.
There is a natural metric on the set $\Sio$ of  infinite words 
over a finite alphabet 
$\Si$ containing at least two letters which is called the {\it prefix metric} and is defined as follows. For $u, v \in \Sio$ and 
$u\neq v$ let $\delta(u, v)=2^{-l_{\mathrm{pref}(u,v)}}$ where $l_{\mathrm{pref}(u,v)}$ 
 is the first integer $n$
such that the $(n+1)^{st}$ letter of $u$ is different from the $(n+1)^{st}$ letter of $v$. 
This metric induces on $\Sio$ the usual  Cantor topology in which the {\it open subsets} of 
$\Sio$ are of the form $W.\Si^\om$, for $W\subseteq \Sis$.
A set $L\subseteq \Si^\om$ is a {\it closed set} iff its complement $\Si^\om - L$ 
is an open set.

For $V \subseteq \Sis$ we denote ${\rm Lim}(V)=\{ x \in \Sio \mid  \exists^\infty  n \geq 1  ~~ x[n] \in V \}$ the set of infinite words over 
$\Si$ having infinitely many prefixes in $V$.  Then the topological closure ${\rm Cl}(L)$ of a set $L\subseteq \Si^\om$ is equal to 
${\rm Lim}({\rm Pref}(L))$. Thus we have also the following characterization of closed subsets of $\Si^\om$:  
a set  $L\subseteq \Si^\om$ is a closed subset of the Cantor space $\Si^\om$ iff  
$L={\rm Lim}({\rm Pref}(L))$.

We   now recall 
the definition of the {\it Borel Hierarchy} of subsets of $X^\om$. 

\begin{defi}
For a non-null countable ordinal $\alpha$, the classes ${\bf \Si}^0_\alpha$
 and ${\bf \Pi}^0_\alpha$ of the Borel Hierarchy on the topological space $X^\om$ 
are defined as follows:
 ${\bf \Si}^0_1$ is the class of open subsets of $X^\om$, 
 ${\bf \Pi}^0_1$ is the class of closed subsets of $X^\om$, 
 and for any countable ordinal $\alpha \geq 2$: 
\nl ${\bf \Si}^0_\alpha$ is the class of countable unions of subsets of $X^\om$ in 
$\bigcup_{\gamma <\alpha}{\bf \Pi}^0_\gamma$.
 \nl ${\bf \Pi}^0_\alpha$ is the class of countable intersections of subsets of $X^\om$ in 
$\bigcup_{\gamma <\alpha}{\bf \Si}^0_\gamma$.

A set $L\subseteq X^\om$ is Borel iff it is in the union $\bigcup_{\alpha < \om_1} {\bf \Si}^0_\alpha = \bigcup_{\alpha < \om_1} {\bf \Pi}^0_\alpha$, where  
$\om_1$ is the first uncountable ordinal. 
\end{defi}

\noi    
There are also some subsets of $X^\om$ which are not Borel. 
In particular 
the class of Borel subsets of $X^\om$ is strictly included into 
the class  ${\bf \Si}^1_1$ of {\it analytic sets} which are 
obtained by projection of Borel sets. The  {\it co-analytic sets}  are the complements of 
analytic sets.  

\begin{defi} 
A subset $A$ of  $X^\om$ is in the class ${\bf \Si}^1_1$ of {\it analytic} sets
iff there exist a finite alphabet $Y$ and a Borel subset $B$  of  $(X \times Y)^\om$ 
such that $ x \in A \lra \exists y \in Y^\om $ such that $(x, y) \in B$, 
where $(x, y)$ is the infinite word over the alphabet $X \times Y$ such that
$(x, y)(i)=(x(i),y(i))$ for each  integer $i\geq 1$.
\end{defi} 

We now recall the notion of  completeness with regard to reduction by continuous functions. 
For a countable ordinal  $\alpha\geq 1$, a set $F\subseteq X^\om$ is said to be 
a ${\bf \Si}^0_\alpha$  
(respectively,  ${\bf \Pi}^0_\alpha$, ${\bf \Si}^1_1$)-{\it complete set} 
iff for any set $E\subseteq Y^\om$  (with $Y$ a finite alphabet): 
 $E\in {\bf \Si}^0_\alpha$ (respectively,  $E\in {\bf \Pi}^0_\alpha$,  $E\in {\bf \Si}^1_1$) 
iff there exists a continuous function $f: Y^\om \ra X^\om$ such that $E = f^{-1}(F)$. 

 We now   recall the definition of classes of the arithmetical hierarchy of $\om$-languages, see \cite{Staiger97}. 
Let $X$ be a finite alphabet. An \ol~ $L\subseteq X^\om$  belongs to the class 
$\Si_n$ if and only if there exists a recursive relation 
$R_L\subseteq (\mathbb{N})^{n-1}\times X^\star$  such that:
\nl $~~~~~~~~~~~~~~~~~ ~~~~~~~~L = \{\sigma \in X^\om \mid \exists a_1\ldots Q_na_n  \quad (a_1,\ldots , a_{n-1}, 
\sigma[a_n+1])\in R_L \},$
\nl where $Q_i$ is one of the quantifiers $\fa$ or $\exists$ 
(not necessarily in an alternating order). An $\om$-language $L\subseteq X^\om$  belongs to the class 
$\Pi_n$ if and only if its complement $X^\om - L$  belongs to the class 
$\Si_n$.  
The  class  $\Si^1_1$ is the class of {\it effective analytic sets} which are 
 obtained by projection of arithmetical sets.
An \ol~ $L\subseteq X^\om$  belongs to the class 
$\Si_1^1$ if and only if there exists a recursive relation 
$R_L\subseteq \mathbb{N}\times \{0, 1\}^\star \times X^\star$  such that:
$$L = \{\sigma \in X^\om  \mid \exists \tau (\tau\in \{0, 1\}^\om \wedge \fa n \exists m 
 ( (n, \tau[m], \sigma[m]) \in R_L )) \}.$$
\noi 
 Then an \ol~ $L\subseteq X^\om$  is in the class $\Si_1^1$ iff it is the projection 
of an \ol~ over the alphabet $X\times \{0, 1\}$ which is in the class $\Pi_2$.  The   class $\Pi_1^1$ of  {\it effective co-analytic sets} 
 is simply the class of complements of effective analytic sets. 

Recall that the (lightface) class $\Si_1^1$ of effective analytic sets is strictly included into the (boldface) class ${\bf \Si}^1_1$ of analytic sets. 

 Recall that a B\"uchi Turing machine is just a Turing machine working on infinite inputs with a B\"uchi-like acceptance condition, and 
that the class of  $\om$-languages accepted by  B\"uchi Turing machines is the class $ \Si^1_1$ of effective analytic sets  \cite{CG78b,Staiger97}.
On the other hand, one can  construct, using a classical  construction (see for instance  \cite{HopcroftMotwaniUllman2001}),  from a  B\"uchi 
Turing machine $\mathcal{T}$,  a $2$-counter  B\"uchi  automaton $\mathcal{A}$ accepting the same $\om$-language. 
Thus one can state the following proposition. 

\begin{Pro}\label{tm}
An \ol~ $L\subseteq X^\om$ is in the class $\Si_1^1$
iff it is accepted by a non deterministic  B\"uchi  Turing machine,  hence  iff it is in the class ${\bf BCL}(2)_\om$. 
\end{Pro}

\section{Context-free Gale-Stewart games}

We first recall the definition of Gale-Stewart games. 

\begin{defi}[\cite{Jech}]
Let $A\subseteq X^\om$, where $X$ is a finite alphabet.
 The Gale-Stewart  game $G(A)$ is a game with perfect
 information between two players. Player 1 first writes a letter
$a_1\in X$, then Player 2 writes a letter $b_1\in X$,
 then Player 1 writes $a_2\in X$, and so on $\ldots$
After $\om$ steps, the two players have composed a word 
$x =a_1b_1a_2b_2\ldots$ of $X^\om$.
 Player 1 wins the play iff $x \in A$, otherwise Player 2
wins the play.

Let $A\subseteq X^\om$ and $G(A)$ be the associated Gale-Stewart  game. A strategy for Player 1 
is a function $F_1: (X^2)^\star \ra X$ and a strategy for Player 2 is a function $F_2: (X^2)^\star X  \ra X$. 
Player 1 follows the strategy $F_1$ in a play if  for each integer $n\geq 1$ ~~  $a_n = F_1(a_1b_1a_2b_2 \cdots a_{n-1}b_{n-1})$. If Player 1 
wins every play in which she has followed the strategy $F_1$, then we say that 
the strategy $F_1$ is a winning strategy (w.s.)  for Player 1.  The notion of winning strategy for Player 2 is defined in a similar manner.  

 The game $G(A)$  is said to be determined if one of the two players has a winning strategy.

 We shall denote {\bf Det}($\mathcal{C}$), where $\mathcal{C}$ is a class of $\om$-languages, 
the sentence : ``Every Gale-Stewart  game $G(A)$, where $A\subseteq X^\om$ is an $\om$-language in the class $\mathcal{C}$, is determined". 
\end{defi}

Notice that, in the whole paper, we assume that ZFC is consistent, and all results, lemmas, propositions, theorems, 
are stated in ZFC unless we explicitely give another axiomatic framework. 

We can now state our first result.

\begin{Pro}\label{the1}
{\bf Det}($\Si_1^1$)  $\Longleftrightarrow$ {\bf Det}({\bf r}-${\bf BCL}(8)_\om$). 
\end{Pro}

\proo   The implication {\bf Det}($\Si_1^1$)  $\Longrightarrow$ {\bf Det}({\bf r}-${\bf BCL}(8)_\om$) is obvious since 
{\bf r}-${\bf BCL}(8)_\om$ $\subseteq \Si_1^1$. 

 To prove the reverse implication, we assume that {\bf Det}({\bf r}-${\bf BCL}(8)_\om$) holds and we are going to show that then 
every Gale-Stewart game $G(A)$, where $A\subseteq X^\om$ is an $\om$-language in the class $\Si_1^1$, or equivalently in the class 
${\bf BCL}(2)_\om$ by Proposition \ref{tm}, is determined. 

 Let then $L \subseteq \Si^\om$, where $\Si$ is a finite alphabet,  be an 
$\om$-language in the class  ${\bf BCL}(2)_\om$. 

 Let   $E$ be a new letter not in 
$\Si$,  $S$ be an integer $\geq 1$, and $\theta_S: \Sio \ra (\Sigma \cup \{E\})^\om$ be the 
function defined, for all  $x \in \Sio$, by: 
$$ \theta_S(x)=x(1).E^{S}.x(2).E^{S^2}.x(3).E^{S^3}.x(4) \ldots 
x(n).E^{S^n}.x(n+1).E^{S^{n+1}} \ldots $$

We proved in \cite{Fin-mscs06} that if    $k=cardinal(\Si)+2$, $S \geq (3k)^3$ is an integer, then one can effectively construct  from 
a    B\"uchi $2$-counter automaton  $\mathcal{A}_1$  accepting $L$   a real time 
B\"uchi $8$-counter automaton $\mathcal{A}_2$ such that $L(\mathcal{A}_2)=\theta_S(L)$. 
In the sequel we assume that we have fixed an integer $S \geq (3k)^3$ which is {\it even}. 

Notice that the set $\theta_S(\Sio)$ is a closed subset of the Cantor space $\Sio$.  An $\om$-word 
$x\in (\Sigma \cup \{E\})^\om$ is in $\theta_S(\Sio)^-$ iff it has one prefix which is not in ${\rm Pref}(\theta_S(\Sio))$. 
Let $L' \subseteq (\Sigma \cup \{E\})^\om$ be the set of $\om$-words $y \in (\Sigma \cup \{E\})^\om$ for which  there is an integer $n\geq 1$ such that 
$y[2n-1]\in {\rm Pref}(\theta_S(\Sio))$ and $y[2n]\notin {\rm Pref}(\theta_S(\Sio))$. 
It is easy to see that $L'$ is  accepted 
by a real time B\"uchi $2$-counter automaton. 

The class 
{\bf r}-${\bf BCL}(8)_\om \supseteq$ {\bf r}-${\bf BCL}(2)_\om$  is closed under  finite union in an effective way, so 
$\theta_S(L) \cup  L'$ is accepted by a real time B\"uchi $8$-counter automaton $\mathcal{A}_3$ which can be effectively constructed 
from     $\mathcal{A}_2$. 

As we have assumed that {\bf Det}({\bf r}-${\bf BCL}(8)_\om$) holds, the game $G(\theta_S(L) \cup  L')$ is determined, i.e. 
one of the two players has a w.s. in the game $G(\theta_S(L) \cup  L')$. We now show that the 
game $G(L)$ is itself determined. 

We shall say that, during  an infinite play, Player 1 ``goes out" of the {\it closed} set 
$\theta_S(\Sio)$ if the final  play $y$ composed by the two players has a prefix $y[2n]\in {\rm Pref}(\theta_S(\Sio))$ such that 
$y[2n+1]\notin {\rm Pref}(\theta_S(\Sio))$. We define in a similar way the sentence  ``Player 2 goes out of the {\it closed} set 
$\theta_S(\Sio)$".

 Assume first that  Player 1 has a w.s.  $F_1$ in the game $G(\theta_S(L) \cup  L')$.    Then Player 1 never ``goes out" of the set 
$\theta_S(\Sio)$ when she follows this w.s. because otherwise the final play $y$ composed by the two players 
has a prefix $y[2n]\in {\rm Pref}(\theta_S(\Sio))$ such that 
$y[2n+1]\notin {\rm Pref}(\theta_S(\Sio))$ and thus $y\notin \theta_S(L) \cup  L'$. 
Consider now a play in which Player 2 does not go out of $\theta_S(\Sio)$.  If player  1 follows her w.s. $F_1$  then the two players remain in the set 
$\theta_S(\Sio)$. But we have fixed $S$ to be an {\bf even} integer. So the two players compose an $\om$-word 
$$\theta_S(x)=x(1).E^{S}.x(2).E^{S^2}.x(3).E^{S^3}.x(4) \ldots 
x(n).E^{S^n}.x(n+1).E^{S^{n+1}} \ldots$$  
and the letters $x(k)$ are written by player 1 for $k$ an odd integer and by Player 2 for $k$ an even integer because $S$ is even. 
Moreover Player 1 wins the play iff the $\om$-word $x(1)x(2)x(3)\ldots 
x(n) \ldots$ is in $L$. This implies that Player 1 has also a w.s. in the game $G(L)$. 

 Assume now that  Player 2 has a w.s.  $F_2$ in the game $G(\theta_S(L) \cup  L')$.     Then Player 2 never ``goes out" of the set 
$\theta_S(\Sio)$ when he follows this w.s. because otherwise the final play $y$ composed by the two players 
has a prefix $y[2n-1]\in {\rm Pref}(\theta_S(\Sio))$ such that 
$y[2n]\notin {\rm Pref}(\theta_S(\Sio))$ and thus $y\in  L'$ hence also $y \in \theta_S(L) \cup  L'$. 
Consider now a play in which Player 1 does not go out of $\theta_S(\Sio)$.  If player  2 follows his w.s. $F_2$  then the two players remain in the set 
$\theta_S(\Sio)$.  So the two players compose an $\om$-word 
$$\theta_S(x)=x(1).E^{S}.x(2).E^{S^2}.x(3).E^{S^3}.x(4) \ldots 
x(n).E^{S^n}.x(n+1).E^{S^{n+1}} \ldots$$  
where the letters $x(k)$ are written by player 1 for $k$ an odd integer and by Player 2 for $k$ an even integer. 
Moreover Player 2 wins the play iff the $\om$-word $x(1)x(2)x(3)\ldots 
x(n) \ldots$ is not in $L$. This implies that Player 2 has also  a w.s. in the game $G(L)$. 
\ep 

\begin{theorem}\label{the2}
{\bf Det}($\Si_1^1$)  $\Longleftrightarrow$     {\bf Det}(${\bf CFL}_\om$)            $\Longleftrightarrow$ {\bf Det}(${\bf BCL}(1)_\om$). 
\end{theorem}

\proo   The implications {\bf Det}($\Si_1^1$) $\Longrightarrow$ {\bf Det}(${\bf CFL}_\om$) 
 $\Longrightarrow$ {\bf Det}(${\bf BCL}(1)_\om$) are obvious since 
${\bf BCL}(1)_\om$ $\subseteq$    ${\bf CFL}_\om$  $\subseteq  \Si_1^1$. 

To prove the reverse implication    {\bf Det}(${\bf BCL}(1)_\om$)   $\Longrightarrow$ {\bf Det}($\Si_1^1$), 
we assume that {\bf Det}(${\bf BCL}(1)_\om$) holds and we are going to show that then 
every Gale-Stewart game $G(L)$, where $L\subseteq X^\om$ is an $\om$-language in the class {\bf r}-${\bf BCL}(8)_\om$ is determined.  
Then Proposition  \ref{the1} will imply that {\bf Det}($\Si_1^1$)  also holds. 
Let then $L(\mathcal{A}) \subseteq \Gao$, where $\Ga$ is a finite alphabet and $\mathcal{A}$ is a real time B\"uchi $8$-counter automaton. 

We now recall  the following  coding  which was used in the paper \cite{Fin-mscs06}. 

Let    $K = 2 \times 3 \times 5 \times 7 \times 11 \times 13 \times 17 \times 19 = 9699690$       be       the product of the eight first prime numbers. 
 An $\om$-word $x\in \Gao$ was coded by the $\om$-word 
$$h_K(x)=A.C^K.x(1).B.C^{K^2}.A.C^{K^2}.x(2).B.C^{K^3}.A.C^{K^3}.x(3).B \ldots  
B.C^{K^n}.A.C^{K^n}.x(n).B \ldots  $$

\noi over the alphabet $\Ga_1=\Ga \cup \{A, B, C\}$, where $A, B, C$ are new letters not in $\Ga$. 
We are going to use here a slightly different coding which we now define.  Let then 

\hs $h(x)=C^K.C.A.x(1).C^{K^2}.A.C^{K^2}.C.x(2).B.C^{K^3}.A.C^{K^3}.C.A.x(3) \ldots  $

\hs ~~~~~~~~~~~~~~~~~~~ $ \ldots C^{K^{2n}}.A.C^{K^{2n}}.C.x(2n).B.C^{K^{2n+1}}.A.C^{K^{2n+1}}.C.A.x(2n+1) \ldots $

\hs \noi  We now explain the rules used to obtain  the $\om$-word $h(x)$ from the $\om$-word $h_K(x)$. 

(1) The first letter $A$ of the word $h_K(x)$ has been suppressed. 

(2) The letters $B$ following a letter $x(2n+1)$, for $n\geq 1$,  have been suppressed. 

(3)  A letter $C$ has been added before each letter $x(2n)$, for $n\geq 1$. 

(4)  A block of two letters $C.A$  has been added before each letter $x(2n+1)$, for $n\geq 1$. 

\noi  The reasons behind this changes are the following ones. Assume that two players 
alternatively write letters from the alphabet $\Ga_1=\Ga \cup \{A, B, C\}$
and that they finally produce an $\om$-word in the form $h(x)$.  
Due to the above changes we have now the two following properties which will be useful in the sequel.

(1) The letters $x(2n+1)$, for $n\geq 0$,  have been  written by Player 1, and the letters $x(2n)$, for $n\geq 1$, have been  written by Player 2. 

(2) After a sequence of consecutive letters $C$, the first letter which is not a C has always been  written by Player 2. 

\noi  We  proved in \cite{Fin-mscs06} that, from a  real time B\"uchi $8$-counter automaton $\mathcal{A}$ accepting $L(\mathcal{A}) \subseteq \Gao$, 
one can effectively construct a  B\"uchi $1$-counter automaton $\mathcal{A}_1$ accepting the $\om$-language 
$h_K( L(\mathcal{A}) )$$ \cup h_K(\Ga^{\om})^-$.
We can easily check that the  changes  in $h_K(x)$ leading to the coding $h(x)$ have no influence with 
 regard to the proof of this result in \cite{Fin-mscs06} and thus  one can also effectively construct 
a  B\"uchi $1$-counter automaton $\mathcal{A}_2$ accepting the $\om$-language 
$h( L(\mathcal{A}) )$$ \cup h(\Ga^{\om})^-$.

 On the other hand we can remark that all $\om$-words  in the form $h(x)$ belong to the $\om$-language $H \subseteq (\Ga_1)^\om$ of $\om$-words $y$ of 
the following form: 

\hs $y=C^{n_1}.C.A.x(1).C^{n_2}.A.C^{n'_2}.C.x(2).B.C^{n_3}.A.C^{n'_3}.C.A.x(3) \ldots  $

\hs ~~~~~~~~~~~~~~~~~~~ $ \ldots C^{n_{2n}}.A.C^{n'_{2n}}.C.x(2n).B.C^{n_{2n+1}}.A.C^{n'_{2n+1}}.C.A.x(2n+1) \ldots $

\hs where for all integers $i\geq 1$ the letters $x(i)$ belong to $\Ga$ and the $n_i$, $n'_i$, are even non-null integers. 

\hs An important fact is the following property of $H$ which extends the same property of  the set $h(\Gao)$. 
Assume that two players 
alternatively write letters from the alphabet $\Ga_1=\Ga \cup \{A, B, C\}$
and that they finally produce an $\om$-word $y$ in $H$ in the above form.  Then we have the two following facts: 

(1)  The letters $x(2n+1)$, for $n\geq 0$,  have been  written by Player 1, and the letters $x(2n)$, for $n\geq 1$, have been  written by Player 2. 

(2)  After a sequence of consecutive letters $C$, the first letter which is not a C has always been  written by Player 2. 

Let now $V={\rm Pref}(H) \cap (\Ga_1)^\star .C$. So a finite word over   the alphabet $\Ga_1$ is in $V$ iff it is a 
prefix  of some word in  $H$ and its  last letter is a $C$. It is easy to see that the topological closure of $H$ is 
$${\rm Cl}(H)=H ~   \cup~    V.C^\om. $$
Notice that an $\om$-word in ${\rm Cl}(H)$ is not in $ h(\Ga^{\om})$ iff a sequence of consecutive letters $C$ has not the good length. Thus if 
two players alternatively write letters from the alphabet $\Ga_1$ and produce an $\om$-word $y \in {\rm Cl}(H) - h(\Ga^{\om})$ then 
it is Player 2 who has gone out of the set $h(\Ga^{\om})$ at some step of the play. This will be important in the sequel. 

\hs  It is very easy to see that the $\om$-language $H$ is regular and to construct a B\"uchi automaton $\mathcal{H}$ accepting it. 
Moreover it is known that the class ${\bf BCL}(1)_\om$ is effectively closed under intersection with regular $\om$-languages 
(this can be seen using a classical construction of a product automaton). Thus 
one can also construct a  B\"uchi $1$-counter automaton $\mathcal{A}_3$ accepting the $\om$-language 
$h( L(\mathcal{A}) )$$ \cup [ h(\Ga^{\om})^- \cap H ]$.

We denote also $U$ the set of finite words $u$ over $\Ga_1$ such that $|u|=2n$ for some integer $n\geq 1$ and $u[2n-1] \in {\rm Pref}(H)$ and 
$u=u[2n] \notin {\rm Pref}(H)$. 

Now we set: 
$$\mathcal{L}~ ~   =  ~ ~  h( L(\mathcal{A}) ) ~ ~  \cup ~ ~   [ h(\Ga^{\om})^- \cap H ] ~ ~   \cup~ ~    V.C^\om ~ ~  \cup ~ ~  U.(\Ga_1)^\om$$
\noi We have already seen that the $\om$-language $h( L(\mathcal{A}) )$$ \cup [ h(\Ga^{\om})^- \cap H ]$ is accepted by a 
B\"uchi $1$-counter automaton $\mathcal{A}_3$. On the other hand the $\om$-language $H$ is regular and it is accepted by a 
B\"uchi automaton $\mathcal{H}$. Thus the finitary language  ${\rm Pref}(H)$ is also regular,   the languages $U$ and $V$ 
are also regular,  and the $\om$-languages $V.C^\om$ and $U.(\Ga_1)^\om$ are regular. This implies that one can construct a 
B\"uchi $1$-counter automaton $\mathcal{A}_4$ accepting the language $\mathcal{L}$. 

\hs By hypothesis  we assume that {\bf Det}(${\bf BCL}(1)_\om$) holds and thus the game $G(\mathcal{L})$ is determined. We are going to show that 
this implies that the game $G( L(\mathcal{A}) )$ itself is determined. 

\hs Assume firstly that Player 1 has a winning strategy $F_1$  in the game $G(\mathcal{L})$.  

If during an infinite play, the two players compose an infinite word $z$, and  Player 2 ``does not go out of the set $h(\Ga^{\om})$" 
then we claim that also Player 1, following her strategy $F_1$, ``does not go out  of the set $h(\Ga^{\om})$". 
Indeed if Player 1 goes out  of the set $h(\Ga^{\om})$ then due to the above remark this would imply that Player 1 also goes out of the set 
${\rm Cl}(H)$: there is an integer $n\geq 0$ such that $z[2n] \in {\rm Pref}(H)$ but  $z[2n+1] \notin {\rm Pref}(H)$. 
So  $z \notin h( L(\mathcal{A}) ) ~   \cup ~   [ h(\Ga^{\om})^- \cap H ]  \cup ~  V.C^\om$.  
Moreover it follows from  the definition of $U$  that $z \notin  U.(\Ga_1)^\om$. 
Thus If Player 1  goes out  of the set $h(\Ga^{\om})$ then she looses the game. 

Consider now  an infinite play in which   Player 2 ``does not go out of the set $h(\Ga^{\om})$".  
Then Player 1, following her strategy $F_1$, ``does not go out  of the set $h(\Ga^{\om})$".  Thus the two players write an infinite word $z=h(x)$ for some 
infinite word $x\in \Ga^{\om}$. But the  letters $x(2n+1)$, for $n\geq 0$,  have been  written by Player 1, and the letters $x(2n)$, for $n\geq 1$, 
have been  written by Player 2.  Player 1 wins the play iff $x\in L(\mathcal{A})$  and Player 1 wins always the play when she uses her strategy $F_1$. 
This implies that Player 1 has also a w.s. in the game  $G( L(\mathcal{A}) )$. 

\hs Assume now that Player 2 has a winning strategy $F_2$  in the game $G(\mathcal{L})$. 
 
If during an infinite play, the two players compose an infinite word $z$, and  Player 1 ``does not go out of the set $h(\Ga^{\om})$" 
then we claim that also Player 2, following his strategy $F_2$, ``does not go out  of the set $h(\Ga^{\om})$". Indeed if 
Player 2 goes out   of the set $h(\Ga^{\om})$ and the final play $z$ remains in ${\rm Cl}(H)$ then 
$z\in  [ h(\Ga^{\om})^- \cap H ] \cup V.C^\om \subseteq \mathcal{L}$ and 
Player 2 looses.  If Player 1 does not go out of the set ${\rm Cl}(H)$ and 
at some step of the play, Player 2 goes out of ${\rm Pref}(H)$, i.e. there is an integer $n\geq 1$ such that $z[2n-1] \in {\rm Pref}(H)$
and $z[2n] \notin {\rm Pref}(H)$,  then $z \in U.(\Ga_1)^\om \subseteq \mathcal{L}$ and Player 2 looses. 

Assume now that Player 1 ``does not go out of the set $h(\Ga^{\om})$". Then  Player 2 follows his w. s.  $F_2$, 
and then ``never goes out  of the set $h(\Ga^{\om})$". 
 Thus the two players write an infinite word $z=h(x)$ for some 
infinite word $x\in \Ga^{\om}$. But the  letters $x(2n+1)$, for $n\geq 0$,  have been  written by Player 1, and the letters $x(2n)$, for $n\geq 1$, 
have been  written by Player 2.  Player 2  wins the play iff $x\notin L(\mathcal{A})$  and Player 2 wins always the play when he uses his strategy $F_2$. 
This implies that Player 2 has also a w.s. in the game  $G( L(\mathcal{A}) )$. 
\ep 

\hs  Looking carefully at the above proof, we can obtain a stronger result: 

\begin{theorem}\label{the3}
{\bf Det}($\Si_1^1$)  $\Longleftrightarrow$     {\bf Det}(${\bf CFL}_\om$)     
$\Longleftrightarrow$ {\bf Det}({\bf r}-${\bf BCL}(1)_\om$). 
\end{theorem}

 \section{Context-free Wadge games}

We first recall the notion of Wadge games. 

\begin{defi}[Wadge \cite{Wadge83}]  Let 
$L\subseteq X^\om$ and $L'\subseteq Y^\om$. 
The Wadge game  $W(L, L')$ is a game with perfect information between two players,
Player 1 who is in charge of $L$ and Player 2 who is in charge of $L'$.
 Player 1 first writes a letter $a_1\in X$, then Player 2 writes a letter
$b_1\in Y$, then Player 1 writes a letter $a_2\in  X$, and so on. 
 The two players alternatively write letters $a_n$ of $X$ for Player 1 and $b_n$ of $Y$
for Player 2.
 After $\om$ steps,  Player 1 has written an $\om$-word $a\in X^\om$ and  Player 2
has written an $\om$-word $b\in Y^\om$.
 Player 2 is allowed to skip, even infinitely often, provided he really writes an
$\om$-word in  $\om$ steps.
Player 2 wins the play iff [$a\in L \lra b\in L'$], i.e. iff: 
~~~~  [($a\in L ~{\rm and} ~ b\in L'$)~ {\rm or} ~ 
($a\notin L ~{\rm and}~ b\notin L'~{\rm and} ~ b~{\rm is~infinite}  $)].
\end{defi}

\noi
Recall that a strategy for Player 1 is a function 
$\sigma :(Y\cup \{s\})^\star\ra X$.
And a strategy for Player 2 is a function $f:X^+\ra Y\cup\{ s\}$.
The strategy  $\sigma$ is a winning strategy  for Player 1 iff she always wins a play when
 she uses the strategy $\sigma$, i.e. when the  $n^{th}$  letter she writes is given
by $a_n=\sigma (b_1\ldots b_{n-1})$, where $b_i$ is the letter written by Player 2 
at step $i$ and $b_i=s$ if Player 2 skips at step $i$.
 A winning strategy for Player 2 is defined in a similar manner.

The game $W(L, L')$ is said to be determined if one of the two players has a winning strategy. 
\noi In the sequel we shall denote {\bf W-Det}($\mathcal{C}$), where $\mathcal{C}$ is a class of $\om$-languages, 
the sentence: ``All Wadge games $W(L, L')$,  where $L\subseteq X^\om$ and  $L'\subseteq Y^\om$ are $\om$-languages 
in the class $\mathcal{C}$, are determined". 

\hs  There is a close relationship between Wadge reducibility
 and games.

\begin{defi}[Wadge \cite{Wadge83}] Let $X$, $Y$ be two finite alphabets. 
For $L\subseteq X^\om$ and $L'\subseteq Y^\om$, $L$ is said to be Wadge reducible to $L'$
($L\leq _W L')$ iff there exists a continuous function $f: X^\om \ra Y^\om$, such that
$L=f^{-1}(L')$.
 $L$ and $L'$ are Wadge equivalent iff $L\leq _W L'$ and $L'\leq _W L$. 
This will be denoted by $L\equiv_W L'$. And we shall say that 
$L<_W L'$ iff $L\leq _W L'$ but not $L'\leq _W L$.

\noi
 The relation $\leq _W $  is reflexive and transitive,
 and $\equiv_W $ is an equivalence relation.
\nl The {\it equivalence classes} of $\equiv_W $ are called {\it Wadge degrees}. 
\end{defi}

\begin{theorem} [Wadge] Let $L\subseteq X^\om$ and $L'\subseteq Y^\om$  where
$X$ and $Y$ are finite  alphabets. Then  $L\leq_W L'$ if and only if  Player 2 has a 
winning strategy  in the Wadge game $W(L, L')$.
\end{theorem}

The Wadge hierarchy $WH$ is the class of Borel subsets of a set  $X^\om$, where  $X$ is a finite set,
equipped with $\leq _W $ and with $\equiv_W $. Using Wadge games, Wadge proved that, up to the complement and $\equiv _W$, 
it is a well ordered hierarchy which 
provides a  great refinement of the Borel hierarchy. 

\hs We can now state  the following  result on determinacy of context-free Wadge games. 

\begin{theorem}\label{thew}
{\bf Det}($\Si_1^1$)  $\Longleftrightarrow$     {\bf W-Det}(${\bf CFL}_\om$)     $\Longleftrightarrow$ {\bf W-Det}(${\bf BCL}(1)_\om$)    
$\Longleftrightarrow$ {\bf W-Det}({\bf r}-${\bf BCL}(1)_\om$). 
\end{theorem}

\noi Recall that, assuming that ZFC is consistent, there are some models of ZFC in which {\bf Det}($\Si_1^1$) does not hold. Therefore there 
are some models of  ZFC in which some Wadge games $W(L(\mathcal{A}), L(\mathcal{B}))$, where $\mathcal{A}$ and $\mathcal{B}$
are B\"uchi $1$-counter automata,  are not determined. We are going to prove that this may be also the case when $\mathcal{B}$ is a 
 B\"uchi automaton (without counter). 
 To prove this, we  use a recent result of  \cite{Fin-ICST} and some results of set theory, so we now briefly recall some notions of set theory and
refer the reader to \cite{Fin-ICST} and to a textbook like \cite{Jech} for more background on set theory. 

\hs  The usual axiomatic system ZFC is 
Zermelo-Fraenkel system ZF   plus the axiom of choice  AC. 
 The axioms of ZFC express some  natural facts that we consider to hold in the universe of sets. 
A model ({\bf V}, $\in)$ of  an arbitrary set of axioms $\mathbb{A}$  is a collection  {\bf V} of sets,  equipped with 
the membership relation $\in$, where ``$x \in y$" means that the set $x$ is an element of the set $y$, which satisfies the axioms of   $\mathbb{A}$. 
We  often say `` the model {\bf V}" instead of "the model ({\bf V}, $\in)$".

We say that two sets $A$ and $B$ have same cardinality iff there is a bijection from $A$ onto $B$ and we denote this  by $A \approx B$. 
The relation $\approx$ is an equivalence relation. 
Using the axiom of choice AC, one can prove that any set $A$ can be well-ordered so  there is an ordinal $\gamma$ such that $A \approx \gamma$. 
In set theory the cardinal of the set $A$ is then formally defined as the smallest such ordinal $\gamma$. 
 The infinite cardinals are usually denoted by
$\aleph_0, \aleph_1, \aleph_2, \ldots , \aleph_\alpha, \ldots$
The continuum hypothesis  CH  says that the first uncountable cardinal $\aleph_1$ is equal to $2^{\aleph_0}$ which is the cardinal of the 
continuum. 

  If  {\bf V} is  a model of ZF and ${\bf L}$ is  the class of  {\it constructible sets} of   {\bf V}, then the class  ${\bf L}$    is a model of  
 ZFC + CH.
Notice that the axiom  V=L, which  means ``every set is constructible",   is consistent with  ZFC  because   ${\bf L}$ is a model of 
ZFC + V=L. 

 Consider now a model {\bf V} of  ZFC  and the class of its constructible sets ${\bf L} \subseteq {\bf V}$ which is another 
model of  ZFC.  It is known that 
the ordinals of {\bf L} are also the ordinals of  {\bf V}, but the cardinals  in  {\bf V}  may be different from the cardinals in {\bf L}. 
 In particular,  the first uncountable cardinal in {\bf L}  is denoted 
 $\aleph_1^{\bf L}$, and it is in fact an ordinal of {\bf V} which is denoted $\om_1^{\bf L}$. 
  It is well-known that in general this ordinal satisfies the inequality 
$\om_1^{\bf L} \leq \om_1$.  In a model {\bf V} of  the axiomatic system  ZFC + V=L the equality $\om_1^{\bf L} = \om_1$ holds, but in 
some other models of  ZFC the inequality may be strict and then $\om_1^{\bf L} < \om_1$. 

The following result was proved in \cite{Fin-ICST}. 

\begin{theorem}\label{the4}
 There exists a real-time $1$-counter B\"uchi automaton $\mathcal{A}$, which can be effectively  constructed, such that the topological complexity of the 
$\om$-language $L(\mathcal{A})$ is not determined by the axiomatic system {\rm ZFC}. Indeed it holds that : 
\begin{enumerate}
\item[(1)] ({\rm ZFC + V=L}). ~~~~~~ The $\om$-language $L(\mathcal{A})$ is an analytic but non-Borel  set. 
\item[(2)] ({\rm ZFC} + $\om_1^{\bf L} < \om_1$).  ~~~~The $\om$-language $L(\mathcal{A})$ is a  ${\bf \Pi}^0_2$-set. 
\end{enumerate}
\end{theorem}

We now state the following new result. To prove it we use in particular  the above Theorem \ref{the4},  the link between Wadge games and 
Wadge reducibility, the ${\bf \Pi}^0_2$-completeness of the regular   $\om$-language  
$(0^\star.1)^\om \subseteq \{0, 1\}^\om$,  the Shoenfield's Absoluteness Theorem, and the notion of  extensions of a model of  {\rm ZFC}. 

\begin{theorem}\footnote{This result has been recently exposed in the Workshops GASICS 2010, Paris, September 2010, and GAMES 2010, Oxford, 
September 2010, but it has never been published.}\label{the5}
 Let $\mathcal{B}$ be a B\"uchi automaton accepting the regular   $\om$-language  
$(0^\star.1)^\om \subseteq \{0, 1\}^\om$. Then one can effectively construct a real-time   $1$-counter 
B\"uchi automaton $\mathcal{A}$ such that:  
\begin{enumerate}
\item[(1)]  ({\rm ZFC} + $\om_1^{\bf L} < \om_1$).  Player 2 has a winning strategy $F$ in the Wadge game $W(L(\mathcal{A}), L(\mathcal{B}))$. 
But $F$ can not be recursive and not even hyperarithmetical. 
\item[(2)]  ({\rm ZFC} + $\om_1^{\bf L} = \om_1$). The Wadge game $W(L(\mathcal{A}), L(\mathcal{B}))$ 
 is not determined. 
\end{enumerate}
\end{theorem}

\begin{Rem}
Every model of  {\rm ZFC} is either a model of ({\rm ZFC} + $\om_1^{\bf L} < \om_1$) or a model of 
({\rm ZFC} + $\om_1^{\bf L} = \om_1$). Thus  there are no models of   {\rm ZFC}  in which 
Player 1  has a 
winning strategy in the Wadge game $W(L(\mathcal{A}), L(\mathcal{B}))$. 
\end{Rem}

\begin{Rem}
In order to prove Theorem \ref{the5} we do not need to use any large cardinal axiom  or even the consistency of such an axiom, like the 
axiom of analytic determinacy. 
\end{Rem}

\section{Concluding remarks}
We have proved that the  determinacy of   Gale-Stewart games whose winning sets are accepted by 
(real-time)  $1$-counter B\"uchi automata  is  equivalent to the determinacy of (effective) analytic Gale-Stewart games which is known to be a large cardinal 
assumption. 

On the other hand we have proved a similar result about the determinacy of  Wadge games. 
We have also obtained an amazing  result, proving that one can effectively construct a  real-time   $1$-counter 
B\"uchi automaton $\mathcal{A}$ and a B\"uchi automaton $\mathcal{B}$  such that  the sentence ``the Wadge game $W(L(\mathcal{A}), L(\mathcal{B}))$ 
is determined" is actually independent from ZFC.  

Notice that  it   is still unknown whether   the determinacy of  Wadge games       $W(L(\mathcal{A}), L(\mathcal{B}))$, where 
$\mathcal{A}$ and   $\mathcal{B}$ are Muller tree automata (reading infinite labelled trees) ,  is provable within ZFC or needs some large cardinal 
assumptions to be proved.

\end{document}